\documentclass[twocolumn,twoside,slac_two]{revtex4}
\usepackage{amsfonts,amsmath,amssymb,bm}
\usepackage{fancyhdr}
\begin{document}

\title{Evolution of Coupled Scalar and Spinor Particles
\\ in Classical Field Theory}

\author{Maxim Dvornikov}
\affiliation{Department of Physics, P.O. Box 35, FIN-40014, University of
Jyv\"{a}skyl\"{a}, Finland}
\email{maxim.dvornikov@phys.jyu.fi}
\affiliation{IZMIRAN, 142190, Troitsk, Moscow region, Russia}
\email{maxdvo@izmiran.ru}
\begin{abstract}
We study the evolution of mixed scalar as well as spinor fields within the context of 
the classical field theory. The initial condition problem is solved and the fields 
distributions, exactly accounting for the initial conditions, are obtained for both 
scalar and spinor fields. In the system of two coupled fields we consider the special 
case of the initial conditions which are rapidly oscillating functions. It is 
demonstrated that the energy densities of the scalar fields and the intensities of 
the spinor fields coincide with the usual transition and survival probabilities of 
neutrino flavor oscillations in vacuum.
\end{abstract}

\maketitle

\thispagestyle{fancy}

\section{INTRODUCTION}

The mixing between different flavor eigenstates of quarks as well as leptons was 
theoretically predicted more than fifty years ago (see 
Refs.~\cite{GelPai55,Pon58eng}). Since then a considerable progress has been achieved 
in both experimental and theoretical studying of elementary particles systems with 
mixing. One should mention the experimental discoveries of $K$ and $B$ mesons 
oscillations (see, e.g., Refs.~\cite{LanBooImpLed56,Alb87}) as well as the recent 
achievements in solar, atmospheric and reactor neutrino experiments (see 
Refs.~\cite{Aha05,Abe06,Ara05}).

The theoretical description of neutrino flavor oscillations as the quantum mechanical 
evolution of two mixed flavor eigenstates was carried out in Ref.~\cite{GriPon69}. 
More detailed quantum mechanical treatment of neutrino flavor oscillations was given 
in Ref.~\cite{Kay81} where the wave packets approach was proposed. The quantum field 
theory was applied to the problem of neutrino flavor oscillations in 
Ref.~\cite{GiuKimLeeLee93} in which the process of oscillations was explained as the 
propagation of neutrino mass eigenstates described by internal lines of a Feynman 
diagram. The nonperturbative quantum field theory effects in neutrino flavor 
oscillations were discussed in Ref.~\cite{BlaVit95} where the authors considered the 
Fock spaces flavor and mass eigenstates and reproduced the Pontecorvo formula for the 
transition probability as well as obtained the corrections to this expression. We 
analyzed neutrino spin and flavor oscillations on the basis of the classical field 
theory in Refs.~\cite{DvoStu02JHEP,Dvo05,Dvo06EPJC}. It was shown that neutrino 
flavor oscillations in vacuum as well as in external fields can be described within 
the framework of the classical field theory. Finally it is worth noticing that many 
other important references on the considered issue are presented in the 
review~\cite{Beu03} devoted to the theory of neutrino flavor oscillations.

In this paper we continue to study neutrino flavor oscillations in vacuum within the 
context of the classical field theory. \emph{Classical particles} are represented 
with help of the \emph{first quantized} fields in our approach. This terminology is 
borrowed from Ref.~\cite{BogShi59}. In Sec.~\ref{SCALAR} we discuss the evolution of 
$N$ coupled scalar fields. We solve the initial condition problem for this system and 
construct the fields distributions. The energy density of each field is calculated. 
It is shown that the obtained expressions are analogous to the formulae used in 
describing neutrino flavor oscillations in vacuum. In Sec.~\ref{SPINOR} the similar 
problem is solved for $N$ coupled spinor fields. We demonstrate that the intensities 
of the fermions are the same as the transition and survival probabilities of neutrino 
flavor oscillations in vacuum. We discuss our results in Sec.~\ref{CONCLUSION}.

\section{EVOLUTION OF SCALAR PARTICLES\label{SCALAR}}

Let us study the evolution of $N$ coupled scalar fields. The
Lagrangian for this system has the form
\begin{equation}\label{LagrphiN}
  \mathcal{L}(\bm{\varphi})=
  \sum_{k=1}^{N}\mathcal{L}_0(\varphi_k)-
  \sum_{\substack{i,k=1 \\ i\neq k}}^{N}
  g_{ik}\varphi_i^\dag\varphi_k,
\end{equation}
where $g_{ik}=g_{ki}^*$ are the coupling constants responsible for vacuum mixing,
$\bm{\varphi}=(\varphi_1,\dots,\varphi_N)$, and
\begin{equation}\label{Lagr0phi}
  \mathcal{L}_0(\varphi_k)=
  \partial_\mu\varphi_k^\dag\partial^\mu\varphi_k-
  \mathfrak{m}_k^2|\varphi_k|^2,
\end{equation}
is the free field Lagrangian for the field
$\varphi_k$, $\mathfrak{m}_k$ is the mass
corresponding to the field $\varphi_k$. In Eq.~\eqref{LagrphiN} we
present the case of the complex fields $\varphi_k$ which can
correspond to electrically charged particles.

Following the results of our previous work~\cite{Dvo05} we set the
initial conditions problem to describe the evolution of our
system~\eqref{LagrphiN} and \eqref{Lagr0phi}. Supposing that the
initial fields distributions
\begin{equation}\label{inicond}
  \varphi_i(\mathbf{r},0)=f_i(\mathbf{r}),
  \quad
  \dot{\varphi}_i(\mathbf{r},0)=g_i(\mathbf{r}),
\end{equation}
are the given functions, one should find the fields distributions
$\varphi_k(\mathbf{r},t)$ for any subsequent moments of time. It
should be noted that it is necessary to set two initial conditions
since the corresponding wave equation -- Klein-Gordon equation --
is the second order differential equation.

In order to solve the initial conditions problem we introduce the
new set of the fields $u_k(\mathbf{r},t)$, which diagonalize the
Lagrangian~\eqref{LagrphiN}. They are usually called the mass
eigenstates. The fields $u_k$ are related to the fields
$\varphi_k$ by the matrix transformation,
\begin{equation}\label{scalarmeigenst}
  \varphi_i(\mathbf{r},t)=
  \sum_{a=1}^{N}M_{ia}u_a(\mathbf{r},t).
\end{equation}
Note that the masses $\mathfrak{m}_i$ of the fields $\varphi_i$
are related to the masses $m_a$ of the fields $u_a$ by the
following formula:
\begin{equation}\label{scalarmass}
  \mathfrak{m}_i^2=\sum_{a=1}^N |M_{ia}|^2 m_a^2,
\end{equation}
which results from Eqs.~\eqref{LagrphiN}, \eqref{Lagr0phi} and 
\eqref{scalarmeigenst}.

Using our previous work~\cite{Dvo05} we obtain the field
distribution of $\varphi_j(\mathbf{r},t)$ which is consistent with
the initial conditions~\eqref{inicond},
\begin{align}\label{phisol3D}
  \varphi_j(\mathbf{r},t)= &
  \sum_{ia=1}^N
  M_{ja} M^{-1}_{ai}
  \notag
  \\
  &
  \times
  \int \mathrm{d}^3\mathbf{r}'
  \big[
  \dot{D}_a(\mathbf{r}-\mathbf{r}',t)f_i(\mathbf{r}')
  \notag
  \\
  & +
  D_a(\mathbf{r}-\mathbf{r}',t)g_i(\mathbf{r}')
  \big],
\end{align}
where
\begin{equation*}
  D_a(\mathbf{r},t)=
  \int
  \frac{\mathrm{d}^3\mathbf{p}}{(2\pi)^3}
  e^{\mathrm{i}\mathbf{p}\mathbf{r}}
  \frac{\sin\mathcal{E}_a t}{\mathcal{E}_a},
\end{equation*}
is the Pauli-Jordan function and $\mathcal{E}_a=\sqrt{p^2+m_a^2}$.

It was demonstrated in Ref.~\cite{Dvo06EPJC} that it was more
convenient to use the momentum representation rather than the
coordinate one. Therefore we rewrite Eq.~\eqref{phisol3D} in the
form
\begin{align}\label{phisol3Dmom}
  \varphi_j(\mathbf{r},t)= &
  \sum_{ia=1}^N
  M_{ja} M^{-1}_{ai}
  \notag
  \\
  &
  \times
  \int \frac{\mathrm{d}^3\mathbf{p}}{(2\pi)^3}
  e^{\mathrm{i}\mathbf{p}\mathbf{r}}
  \big[
  \dot{D}_a(\mathbf{p},t)f_i(\mathbf{p})
  \notag
  \\
  & +
  D_a(\mathbf{p},t)g_i(\mathbf{p})
  \big],
\end{align}
with help of the Fourier transforms of the Pauli-Jordan function,
\begin{equation}\label{FTPJfunc4D}
  \dot{D}_a(\mathbf{p},t)=\cos\mathcal{E}_a t,
  \quad
  D_k(\mathbf{p},t)=\frac{\sin\mathcal{E}_a t}{\mathcal{E}_a},
\end{equation}
and the initial conditions,
\begin{align}\label{FTinicond}
  f_i(\mathbf{p}) & =\int\mathrm{d}^3\mathbf{r}
  e^{-\mathrm{i}\mathbf{p}\mathbf{r}}f_i(\mathbf{r}),
  \notag
  \\
  g_i(\mathbf{p}) & =\int\mathrm{d}^3\mathbf{r}
  e^{-\mathrm{i}\mathbf{p}\mathbf{r}}g_i(\mathbf{r}).
\end{align}

It is interesting to demonstrate that the total energy is conserved in our system. 
Indeed, the total energy is given by the formula
\begin{equation*}
  \mathbb{E}(t)=
  \int\mathrm{d}^3\mathbf{r}T^{00}_\mathrm{total}(\mathbf{r},t),
\end{equation*}
where $T^{00}_\mathrm{total}$ is the time component of the energy-momentum tensor. 
Using the general expression for $T^{\mu\nu}_\mathrm{total}$,
\begin{equation*}
  T^{\mu\nu}_\mathrm{total}=\eta^{\nu\lambda}\sum_{k=1}^N
  \left(
    \frac{\partial\mathcal{L}}{\partial(\partial_\mu \varphi_k)}
	\partial_\lambda \varphi_k+
	\text{h.\thinspace c.}
  \right)-
  \mathcal{L}\eta^{\mu\nu},
\end{equation*}
where the Lagrangian $\mathcal{L}$ is presented in Eq.~\eqref{LagrphiN}, we receive 
the formula for $\mathbb{E}(t)$ in the form
\begin{align}\label{totent}
  \mathbb{E}(t)= &
  \int\mathrm{d}^3\mathbf{r}
  \bigg\{
    \sum_{k=1}^N
    (|\dot{\varphi}_k(\mathbf{r},t)|^2+
	|\bm{\nabla}\varphi_k(\mathbf{r},t)|^2
	\\
	\notag
	& +
    \mathfrak{m}_k^2|\varphi_k(\mathbf{r},t)|^2)+	
    \sum_{\substack{i,k=1 \\ i\neq k}}^{N}
    g_{ik}\varphi_i^\dag(\mathbf{r},t)\varphi_k(\mathbf{r},t)
  \bigg\}.
\end{align}
The initial value of the total energy can be obtained with help of 
Eqs.~\eqref{inicond}, \eqref{FTinicond} and~\eqref{totent},
\begin{align}\label{FTtoten0}
  \mathbb{E}(0)= &
  \int\frac{\mathrm{d}^3\mathbf{p}}{(2\pi)^3}
  \bigg\{
    \sum_{k=1}^N
    (|g_k(\mathbf{p})|^2+\mathbf{p}^2|f_k(\mathbf{p})|^2
	\notag
	\\
	& +
    \mathfrak{m}_k^2|f_k(\mathbf{p})|^2)+	
    \sum_{\substack{i,k=1 \\ i\neq k}}^{N}
    g_{ik}f_i^\dag(\mathbf{p}) f_k(\mathbf{p})
  \bigg\}.
\end{align}
Using the explicit form for the field distribution $\varphi_j(\mathbf{r},t)$ 
[Eq.~\eqref{phisol3Dmom}] and the Fourier transforms of the Pauli-Jordan function 
[Eq.~\eqref{FTPJfunc4D}] one can rewrite Eq.~\eqref{totent} in the following way:
\begin{align}\label{totentint}
  \mathbb{E}(t)= &
  \sum_{ika=1}^N
  M_{ia} M^{-1}_{ak}
  \int\frac{\mathrm{d}^3\mathbf{p}}{(2\pi)^3}
  \notag
  \\
  & \times
  \left[
    g_i^\dag(\mathbf{p}) g_k(\mathbf{p})+
	\mathcal{E}_a^2 f_i^\dag(\mathbf{p}) f_k(\mathbf{p})
  \right].
\end{align}
In deriving Eq.~\eqref{totentint} we use the unitarity of the matrix $M_{ja}$, i.e. 
$(M^{-1})_{ai}=M^{*}_{ia}$, and the property of the masses eigenstates $u_a$,
\begin{equation*}
  m_a^2\delta_{ab}=
  \sum_{k=1}^N
  \mathfrak{m}_k^2 M_{ka}^{*}M_{kb}+
  \sum_{\substack{i,k=1 \\ i\neq k}}^{N}
  g_{ik}M_{ia}^{*}M_{kb}.
\end{equation*}
With help of the identity
\begin{equation*}
  \sum_{a=1}^N M_{ia} M_{ak}^{-1} m_a^2=
  \mathfrak{m}_i^2\delta_{ik}+g_{ik},
  \quad
  g_{ii}=0,
\end{equation*}
which generalizes Eq.~\eqref{scalarmass} we can transform Eq.~\eqref{totentint} to 
the form which coincides with Eq.~\eqref{FTtoten0}, i.e. we demonstrate that 
$\mathbb{E}(t)=\mathbb{E}(0)$. This explicit calculation shows that the total energy 
in the system of $N$ coupled scalar fields is conserved independently of initial 
conditions.

It is very difficult to analyze Eq.~\eqref{phisol3Dmom} for arbitrary
functions $f_i(\mathbf{p})$ and $g_i(\mathbf{p})$. Thus we pick
out the special form of the initial conditions (see also
Refs.~\cite{Dvo05,Dvo06EPJC}),
\begin{equation}\label{inicondspec}
  f_i(\mathbf{r})=\frac{A_i}{\sqrt{\mathfrak{E}_i}}
  e^{\mathrm{i}\bm{\omega}\mathbf{r}},
  \quad
  g_i(\mathbf{r})=0,
\end{equation}
where $A_i$ is the "amplitude" of the function $f_i$,
$\mathfrak{E}_i=\sqrt{\omega^2+\mathfrak{m}_i^2}$ and
$\bm{\omega}$ are the initial energy and momentum of the field
$\varphi_i$. Note that $1/\sqrt{\mathfrak{E}_i}$ is the
normalization factor; its value will be clarified below [see
Eq.~\eqref{testenergy}]. Using Eq.~\eqref{FTinicond} we obtain the
Fourier transforms of the initial conditions,
\begin{equation}\label{FTinicondspec}
  f_i(\mathbf{p})=\frac{A_i}{\sqrt{\mathfrak{E}_i}}
  (2\pi)^3 \delta^3(\bm{\omega}-\mathbf{p}),
  \quad
  g_i(\mathbf{p})=0.
\end{equation}

The physical quantity measured in an experiment is \emph{not} a
field distribution. It was demonstrated above that the total energy of the system is 
conserved. Therefore we assume that one can detect the energy
density of a \emph{single} scalar field. Constructing the energy-momentum tensor
for the \emph{single} field $\varphi_i$ and taking its time component we obtain
for the energy density,
\begin{align}\label{edgen}
  \mathfrak{H}_i(\mathbf{r},t)= & T^{00}[\varphi_i(\mathbf{r},t)]
  \notag
  \\
  = &
  |\dot{\varphi}_i|^2+|\bm{\nabla}\varphi_i|^2+
  \mathfrak{m}_i^2|\varphi_i|^2.
\end{align}
Note that here we assume that there would be no mixing between different fields 
$\varphi_i$. It can be verified by means of direct calculations that the
following expression:
\begin{equation}\label{testenergy}
  \mathfrak{H}_i(\mathbf{r},0)=|A_i|^2\mathfrak{E}_i,
\end{equation}
results from Eqs.~\eqref{inicondspec} and \eqref{edgen}. Therefore
the normalization factor in Eq.~\eqref{inicondspec} was chosen to
get the correct value of the initial energy density.

Without losing generality we can discuss now the evolution of two
scalar fields. In this simple case the matrix $M_{ja}$ can be
parameterized with help of only one mixing angle $\theta$,
\begin{equation}\label{Mtheta}
  M_{ja}=
  \begin{pmatrix}
    \cos\theta & -\sin\theta \\
    \sin\theta & \cos\theta
  \end{pmatrix}.
\end{equation}
It is also necessary to fix the "amplitudes" $A_i$. They can be
chosen in following form, $A_1=0$ and $A_2=1$. This choice of the
"amplitudes" has rather simple physical meaning: the first field
is absent initially and the second one has the plane wave field
distribution.

Using Eqs.~\eqref{phisol3Dmom}, \eqref{FTinicondspec} and
\eqref{Mtheta} we get the field distributions of $\varphi_{1,2}$:
\begin{align}\label{phispecgen}
  \varphi_1(\mathbf{r},t)= & 
  \frac{1}{\sqrt{\mathfrak{E}_2}}\cos\theta\sin\theta
  e^{\mathrm{i}\bm{\omega}\mathbf{r}}
  \\
  & \times
  (\cos[\mathcal{E}_1(\omega) t]-\cos[\mathcal{E}_2(\omega) t]),
  \notag
  \\
  \varphi_2(\mathbf{r},t)= &
  \frac{1}{\sqrt{\mathfrak{E}_2}}
  e^{\mathrm{i}\bm{\omega}\mathbf{r}}
  \notag
  \\
  & \times
  (\sin^2\theta\cos[\mathcal{E}_1(\omega) t]+
  \cos^2\theta\cos[\mathcal{E}_2(\omega) t]),
  \notag
\end{align}
where $\mathcal{E}_a(\omega)=\sqrt{\omega^2+m_a^2}$. To analyze
the obtained expressions we consider the limiting case of high
initial frequency, $\omega\gg m_{1,2}$, which corresponds to
ultrarelativistic particles.

With help of Eqs.~\eqref{edgen} and \eqref{phispecgen} as well as
in the high frequency approximation we obtain the expressions for
the energy densities of the fields $\varphi_{1,2}$ in the
following form:
\begin{align}\label{edspec}
  \mathfrak{H}_1(t)= &
  \omega
  \left\{
    \sin^2(2\theta)\sin^2[\Delta(\omega)t]+
    \mathcal{O}
    \left(
      \frac{m_a^2}{\omega^2}
    \right)
  \right\},
  \\
  \mathfrak{H}_2(t)= &
  \omega
  \left\{
    1-
    \sin^2(2\theta)\sin^2[\Delta(\omega)t]+
    \mathcal{O}
    \left(
      \frac{m_a^2}{\omega^2}
    \right)
  \right\},
  \notag
\end{align}
where
\begin{equation*}
  \Delta(\omega)=
  \frac{\mathcal{E}_1(\omega)-\mathcal{E}_2(\omega)}{2}\to
  \frac{\Delta m^2}{4\omega}+
  \mathcal{O}
  \left(
    \frac{m_a^2}{\omega^2}
  \right).
\end{equation*}
Here we use the common notation $\Delta m^2=m_1^2-m_2^2$.

Now we can introduce the probability which corresponds to the
transitions from the second eigenstate $\varphi_2$ to the
eigenstate $\varphi_k$ with $k=1,2$: $P^{(\mathrm{scalar})}_{2\to
k}(t)=\mathfrak{H}_k(t)/\omega$. Using Eqs.~\eqref{edspec} we obtain for
the probabilities,
\begin{align}
  \label{scalprtr}
  P^{(\mathrm{scalar})}_{2\to 1}(t)= &
  \sin^2(2\theta)\sin^2
  \left(
    \frac{\Delta m^2}{4\omega}t
  \right),
  \\
  \label{scalprsur}
  P^{(\mathrm{scalar})}_{2\to 2}(t)= &
  1-\sin^2(2\theta)\sin^2
  \left(
    \frac{\Delta m^2}{4\omega}t
  \right).
\end{align}
The function $P^{(\mathrm{scalar})}_{2\to 1}(t)$ in
Eq.~\eqref{scalprtr} is usually called the transition probability
whereas the function $P^{(\mathrm{scalar})}_{2\to 2}$ in
Eq.~\eqref{scalprsur} -- the survival probability. It should be
noted that Eqs.~\eqref{scalprtr} and \eqref{scalprsur} reproduce
the common quantum mechanical expressions for transition and
survival probabilities of neutrino flavor oscillations in vacuum.

\section{EVOLUTION OF SPINOR PARTICLES\label{SPINOR}}

Now let us discuss the evolution of $N$ coupled spinor fields. The
Lagrangian for this system has the form
\begin{equation*}
  \mathcal{L}(\bm{\nu})=
  \sum_{k=1}^{N}\mathcal{L}_0(\nu_k)-
  \sum_{\substack{i,k=1 \\ i\neq k}}^{N}
  g_{ik}\bar{\nu}_i\nu_k,
\end{equation*}
where $\bm{\nu}=(\nu_1,\dots,\nu_N)$ and
\begin{equation*}
  \mathcal{L}_0(\nu_k)=
  \bar{\nu}_k(\mathrm{i}\gamma^\mu\partial_\mu-\mathfrak{m}_k)\nu_k,
\end{equation*}
is the free field Lagrangian for $\nu_k$ and $\mathfrak{m}_k$ are the
masses of $\nu_k$.

Analogously to Sec.~\ref{SCALAR} (see also
Refs.~\cite{Dvo05,Dvo06EPJC}) we set the initial condition
problem for the considered system,
\begin{equation}\label{inicondspin}
  \nu_i(\mathbf{r},0)=\xi_i(\mathbf{r}),
\end{equation}
where $\xi_i(\mathbf{r})$ are the given functions. It should be
mentioned that we have to set only one initial condition in
Eq.~\eqref{inicondspin} since the Dirac equation is the first
order differential equation. Now it is necessary to find the
fields distributions for $\nu_i(\mathbf{r},t)$ at $t>0$ which
would be consistent with Eq.~\eqref{inicondspin}.

With help of Refs.~\cite{Dvo05,Dvo06EPJC} we obtain for
$\nu_j(\mathbf{r},t)$ the following expression:
\begin{align}
  \nu_j(\mathbf{r},t)= &
  \sum_{ia=1}^N
  M_{ja} M^{-1}_{ai}
  \notag
  \\
  &
  \label{nusolviaxi}
  \times
  \int \mathrm{d}^3\mathbf{r}'
  S_a(\mathbf{r}'-\mathbf{r},t)(-\mathrm{i}\gamma^0)\xi_i(\mathbf{r}'),
\end{align}
where
\begin{equation*}
  S_a(\mathbf{r},t)=(\mathrm{i}\gamma^\mu\partial_\mu+m_a)D_a(\mathbf{r},t),
  \quad
  x^\mu=(t,\mathbf{r}),
\end{equation*}
is the Pauli-Jordan function for a spinor field and $m_a$ are the
masses of mass eigenstates which are related to $\mathfrak{m}_i$
by the formula,
\begin{equation*}
  \mathfrak{m}_i=\sum_{a=1}^N |M_{ia}|^2 m_a.
\end{equation*}
It should be noted that Eq.~\eqref{nusolviaxi} is in agreement
with the initial conditions in Eq.~\eqref{inicondspin}.

As it was demonstrated in Sec.~\ref{SCALAR} we have to rewrite
Eq.~\eqref{nusolviaxi} in the momentum representation in order to
avoid computational difficulties. Finally we get
\begin{align}\label{nusolviaxiFT}
  \nu_j(\mathbf{r},t)= &
  \sum_{ia=1}^N
  M_{ja} M^{-1}_{ai}
  \notag
  \\
  &
  \times
  \int \frac{\mathrm{d}^3\mathbf{p}}{(2\pi)^3}
  e^{\mathrm{i}\mathbf{p}\mathbf{r}}
  S_a(-\mathbf{p},t)(-\mathrm{i}\gamma^0)\xi_i(\mathbf{p}),
\end{align}
where
\begin{align}\label{PJpspin}
  S_a(-\mathbf{p},t)= &
  \bigg[
    \cos\mathcal{E}_a t
	\notag
	\\
	& -
    \mathrm{i}\frac{\sin\mathcal{E}_a t}{\mathcal{E}_a}
    (\bm{\alpha}\mathbf{p}+\beta m_a)
  \bigg](i\beta),
\end{align}
and
\begin{equation*}
  \xi_i(\mathbf{p})=\int\mathrm{d}^3\mathbf{r}
  e^{-\mathrm{i}\mathbf{p}\mathbf{r}}\xi_i(\mathbf{r}),
\end{equation*}
are the Fourier transforms of the Pauli-Jordan function and the
initial conditions. Here we adopt the common notation for the
Dirac matrices $\bm{\alpha}=\gamma^0\bm{\gamma}$ and
$\beta=\gamma^0$.

It is interesting to verify that in the classical field theory approach the total 
intensity of all the fields $\nu_k$ is conserved. We assume that
\begin{align}\label{intotprob}
  \mathfrak{I}(0)= & \sum_{i=1}^{N}\int\mathrm{d}^3\mathbf{r}
  |\xi_i(\mathbf{r})|^2
  \notag
  \\
  = &
  \sum_{i=1}^{N}\int
  \frac{\mathrm{d}^3\mathbf{p}}{(2\pi)^3}
  |\xi_i(\mathbf{p})|^2,
\end{align}
has the given value.
Now we are interested in the following quantity:
\begin{equation*}
  \mathfrak{I}(t)=\sum_{j=1}^{N}\int\mathrm{d}^3\mathbf{r}
  |\nu_j(\mathbf{r},t)|^2.
\end{equation*}
With help of Eq.~\eqref{nusolviaxiFT} this expression can be rewritten in the form
\begin{align}\label{int1totprob}
  \mathfrak{I}(t)= & \sum_{jikab=1}^{N}
  [M_{ja} M^{-1}_{ai}]^{*}M_{jb} M^{-1}_{bk}
  \notag
  \\
  & \times    
  \int
  \frac{\mathrm{d}^3\mathbf{p}}{(2\pi)^3}
  \frac{\mathrm{d}^3\mathbf{q}}{(2\pi)^3}
  \mathrm{d}^3\mathbf{r} \thinspace
  e^{\mathrm{i}(\mathbf{q}-\mathbf{p})\mathbf{r}}
  \notag
  \\
  & \times
  \xi_i^\dagger(\mathbf{p})
  [S_a(-\mathbf{p},t)(-\mathrm{i}\gamma^0)]^\dagger
  \notag
  \\
  & \times
  S_b(-\mathbf{q},t)(-\mathrm{i}\gamma^0)
  \xi_k(\mathbf{q}).
\end{align}
Again using the fact that the matrix $M_{ja}$ is the unitary one, we obtain for 
Eq.~\eqref{int1totprob},
\begin{align}\label{int2totprob}
  \mathfrak{I}(t)
  = & \sum_{ika=1}^{N}
  M_{ia} M^{-1}_{ak}
  \notag
  \\
  & \times
  \int
  \frac{\mathrm{d}^3\mathbf{p}}{(2\pi)^3}
  \xi_i^\dagger(\mathbf{p})
  [S_a(-\mathbf{p},t)(-\mathrm{i}\gamma^0)]^\dagger
  \notag
  \\
  & \times
  S_a(-\mathbf{p},t)(-\mathrm{i}\gamma^0)
  \xi_k(\mathbf{p}).
\end{align}
With help of Eq.~\eqref{PJpspin} one can show that
\begin{equation*}
  [S_a(-\mathbf{p},t)(-\mathrm{i}\gamma^0)]^\dagger
  S_a(-\mathbf{p},t)(-\mathrm{i}\gamma^0)=1.
\end{equation*}
Therefore, taking into account Eq.~\eqref{intotprob}, we represent  
Eq.~\eqref{int2totprob} in the form
\begin{equation}\label{fin2totprob}
  \mathfrak{I}(t)=
  \sum_{i=1}^{N}\int
  \frac{\mathrm{d}^3\mathbf{p}}{(2\pi)^3}
  |\xi_i(\mathbf{p})|^2=\mathfrak{I}(0),
\end{equation}
which shows that the the total intensity is conserved.

In the following we consider the evolution of only two coupled
spinor fields to make the results more illustrative. Thus we
assume that the initial conditions are: $\xi_1(\mathbf{r})=0$ and
$\xi_2(\mathbf{r})=e^{\mathrm{i}\bm{\omega}\mathbf{r}}\xi_0$. The discussion of such 
a choice of initial conditions is given in Sec.~\ref{SCALAR}. The mixing
matrix $M_{ja}$ can be also represented in the same form as in
Eq.~\eqref{Mtheta}. Using general Eq.~\eqref{nusolviaxiFT} one
obtains for the fields distributions of $\nu_{1,2}$:
\begin{align}\label{nudistrgen}
  \nu_1(\mathbf{r},t)= & 
  e^{\mathrm{i}\bm{\omega}\mathbf{r}} \sin\theta\cos\theta
  \bigg\{
    \cos[\mathcal{E}_1(\omega) t]-
    \cos[\mathcal{E}_2(\omega) t]
    \notag
    \\
    \notag
    & -
    \mathrm{i}\frac{\sin[\mathcal{E}_1(\omega) t]}{\mathcal{E}_1(\omega)}
    (\bm{\alpha}\bm{\omega}+\beta m_1)
    \\
    \notag
    & +
    \mathrm{i}\frac{\sin[\mathcal{E}_2(\omega) t]}{\mathcal{E}_2(\omega)}
    (\bm{\alpha}\bm{\omega}+\beta m_2)
  \bigg\}\xi_0,
  \\
  \notag
  \nu_2(\mathbf{r},t)= &
  e^{\mathrm{i}\bm{\omega}\mathbf{r}}
  \bigg\{
    \sin^2\theta\cos[\mathcal{E}_1(\omega) t]+
    \cos^2\theta\cos[\mathcal{E}_2(\omega) t]
    \\
    \notag
    &
    -\mathrm{i}\sin^2\theta
    \frac{\sin[\mathcal{E}_1(\omega) t]}{\mathcal{E}_1(\omega)}
    (\bm{\alpha}\bm{\omega}+\beta m_1)
    \\
    & -
    \mathrm{i}\cos^2\theta
    \frac{\sin[\mathcal{E}_2(\omega) t]}{\mathcal{E}_2(\omega)}
    (\bm{\alpha}\bm{\omega}+\beta m_2)
  \bigg\}\xi_0.
\end{align}
The most interesting case is the high frequency approximation of
the initial conditions given in Eq.~\eqref{inicondspin}, i.e. $\omega\gg
m_{1,2}$. Therefore, Eqs.~\eqref{nudistrgen} can be rewritten in
the form
\begin{align}\label{nudistrhf}
  \nu_1(\mathbf{r},t)= & -
  e^{\mathrm{i}\bm{\omega}\mathbf{r}} \sin2\theta
  \sin[\Delta(\omega)t]
  \notag
  \\
  \notag
  & \times
  \left\{
    \sin[\sigma(\omega) t]+
    \mathrm{i}(\bm{\alpha}\mathbf{n})\cos[\sigma(\omega) t]
  \right\}\xi_0
  \\
  \notag
  & +\mathcal{O}(m_a/\omega),
  \\
  \notag
  \nu_2(\mathbf{r},t)= &
  e^{\mathrm{i}\bm{\omega}\mathbf{r}}
  \left\{
    \sin[\Delta(\omega)t]\cos2\theta-
    \mathrm{i}(\bm{\alpha}\mathbf{n})\cos[\Delta(\omega) t]
  \right\}
  \\
  \notag
  & \times
  \left\{
    \sin[\sigma(\omega) t]+
    \mathrm{i}(\bm{\alpha}\mathbf{n})\cos[\sigma(\omega) t]
  \right\}\xi_0
  \\
  & +\mathcal{O}(m_a/\omega),
\end{align}
where
\begin{equation*}
  \sigma(\omega)=\frac{\mathcal{E}_1(\omega)+\mathcal{E}_2(\omega)}{2}
  \to\omega+\frac{m^2_1+m^2_2}{4\omega}+
  \mathcal{O}
  \left(
    \frac{m_a^2}{\omega^2}
  \right).
\end{equation*}

We showed above that the total intensity of all the fields $\nu_k$, which is 
proportional to a squared fields distributions, is conserved in the considered 
system. Therefore we suppose that the measurable quantity of a classical spinor 
particle is the
intensity, $I_k(t)=|\nu_k(\mathbf{r},t)|^2$. Analogously to Sec.~\ref{SCALAR}
we introduce the probabilities for transitions from the second
eigenstate to the eigenstate $k=1,2$ as
$P^{(\mathrm{spinor})}_{2\to k}(t)=I_k(t)$. Using
Eqs.~\eqref{nudistrhf} we express the probabilities in the
following way:
\begin{align}
  \label{traprobspin}
  P^{(\mathrm{spinor})}_{2\to 1}(t)= &
  \sin^2(2\theta)\sin^2
  \left(
    \frac{\Delta m^2}{4\omega}t
  \right),
  \\
  \label{surprobspin}
  P^{(\mathrm{spinor})}_{2\to 2}(t)= &
  1-\sin^2(2\theta)\sin^2
  \left(
    \frac{\Delta m^2}{4\omega}t
  \right).
\end{align}
One can observe that these expressions for
transition~\eqref{traprobspin} and survival~\eqref{surprobspin}
probabilities coincide with the analogous Eqs.~\eqref{scalprtr}
and \eqref{scalprsur} for the coupled scalar fields. Note that 
Eqs.~\eqref{traprobspin} and~\eqref{surprobspin} are consistent with 
Eq.~\eqref{fin2totprob}. The obtained
probabilities formulas are the same as in the quantum mechanical
treatment of neutrino flavor oscillations in vacuum.

\section{CONCLUSION\label{CONCLUSION}}

In conclusion we note that the evolution of coupled scalar as well
as spinor fields has been described in frames of classical field
theory. First we have considered the system of $N$ arbitrary
coupled complex scalar fields and solved the initial conditions
problem for this system. The general fields distributions
consistent with the initial conditions have been found. Note that
these expressions exactly took into account the Lorentz invariance
and were valid in (3+1)-dimensional space-time. The energy conservation law has been 
analyzed for the considered system. Then we have
discussed the simple case of two coupled fields and have
constructed the energy densities of each scalar field. It has been
demonstrated that in ultrarelativistic limit these expressions
were analogous to the transition and the survival probabilities of
neutrino flavor oscillations in vacuum. The calculations in the
present paper have refined the results of our previous
work~\cite{Dvo05} where analogous problem was analyzed in
(1+1)-dimensional space-time. 

The case of $N$ coupled classical
spinor fields has been also considered in the present work. We
have formulated the initial condition problem and constructed the general
expressions for the fields distributions which were valid in
(3+1)-dimensional space-time and were Lorentz invariant. The total intensity of all 
coupled spinor fields has been computed. It has been demonstrated that this quantity 
was conserved independently of the initial conditions. Then we have
discussed the case of rapidly oscillating initial conditions which
corresponded to ultrarelativistic particles. In this case as well as
for only two coupled fermions we have obtained the intensities of
the fields in question. The expressions for the intensities were
also revealed to coincide with transition and survival
probabilities of neutrino flavor oscillations in vacuum. It should
be mentioned that transition
and survival probabilities for coupled classical spinor fields turned out to be the 
same as the analogous formulae for scalar fields.

\bigskip
\begin{acknowledgments}
This research was supported by the Academy of Finland under the 
contract No.~108875 and by a grant of Russian Science Support
Foundation. The author is indebted to the organizers of the IPM
School \& Conference on Lepton and Hadron Physics for the
invitation to participate in this activity.
\end{acknowledgments}

\end{document}